\documentclass[pra,groupedaddress,showpacs]{revtex4}
\usepackage{graphicx}
\begin{document}
    
\title{Improved Semiclassical Approximation for Bose-Einstein Condensates: 
Application to a BEC in an Optical Potential}

\author{Y.B. Band}
\author{I. Towers}
\altaffiliation{Department of Interdisciplinary Studies, Faculty of 
Engineering, Tel-Aviv University, Tel-Aviv 69978, Israel}
\affiliation{Department of Chemistry, Ben-Gurion University of the Negev, 
84105 Beer-Sheva, Israel}
\author{B.A. Malomed}
\affiliation{Department of Interdisciplinary Studies, Faculty of 
Engineering, Tel Aviv University, Tel Aviv 69978, Israel}

\begin{abstract} 
We present semiclassical descriptions of Bose-Einstein condensates for 
configurations with spatial symmetry, e.g., cylindrical symmetry, and 
without any symmetry. The description of the cylindrical case is 
quasi-one-dimensional (Q1D), in the sense that one only needs to solve an 
effective 1D nonlinear Schr\"{o}dinger equation, but the solution 
incorporates correct 3D aspects of the problem. The solution in classically 
allowed regions is matched onto that in classically forbidden regions by a 
connection formula that properly accounts for the nonlinear mean-field 
interaction. Special cases for vortex solutions are treated too. Comparisons 
of the Q1D solution with full 3D and Thomas-Fermi ones are presented. 
\end{abstract} 

\maketitle

\section{Introduction} 
 
Simple approximations for describing Bose-Einstein condensates (BECs)
have been very useful for understanding their physics.  For example,
in the mean-field approximation, in regions where the local density is
large enough, so that the mean-field (nonlinear) term in the
Gross-Pitaevskii equation (GPE) is much larger than the kinetic-energy
term, the Thomas-Fermi (TF) approximation offers such a simplified
description for the ground state of a BEC in a stationary potential
\cite{Dalfovo}.  However, in classically forbidden regions of the
coordinate space, the density is low, and the TF approximation is
invalid.  It is necessary to match the TF approximation in the region
of high density to a description valid near the boundaries of the
classically allowed motion and in the classically forbidden region for
a given external potential.  For dynamic situations, some simple
approximations exist for time-dependent harmonic potentials
\cite{Castin-Dum,Japha-Band}.  It would be very useful to have a
simple effective one-dimensional (1D) approximation that properly
accounts for the 3D character of a BEC, for both static and dynamic
problems in configurations with spatial symmetry, such as a BEC in a
cylindrically symmetric potential (e.g., a harmonic trap with
cylindrical symmetry, with or without an optical potential that varies
in space along the symmetry axis of the harmonic potential, see a
detailed formulation of the model below in Sec.~\ref{S_Model}).  Pedri
\textit{et al.} developed a treatment of this kind \cite{Pedri}. 
Another contribution was made in Ref.~\cite{Salasnich}, which aimed at
a derivation of an effectively one-dimensional (1D) GPE relevant for
the description of 3D BEC by means of the variational approximation. 
The 1D wave function derived in \cite{Salasnich} was defined so that
it has the same density distribution along the symmetry axis of the
system as that which could be obtained by integrating the distribution
produced by the full 3D wave function in the transverse plane.  As a
result, the effective 1D equation derived in Ref.~\cite{Salasnich} had
a non-polynomial nonlinearity.  Moreover, the transverse Gaussian
distribution adopted in Ref.~\cite{Salasnich} to derive the 1D
equation, strongly differs from the actual transverse distribution in
the case of large BEC density (which is well approximated by the TF
form, see below).
 
Here, our objective is to improve upon the treatments of
Refs.~\cite{Pedri,Salasnich} in a number of ways.  In our treatment in
Sec.~\ref {S_Semicl}, regions of physical space in which the density
is sufficient for the application of the mean-field approximation, and
those in which the density is low, are treated completely differently. 
In Sec.~\ref{S_Matching_Conditions} we develop a connection
(crossover) formula for the wave function in these regions, in analogy
with the commonly known formulas between classically allowed and
forbidden regions in ordinary quantum mechanics \cite{LLQM} (the
difference from the connection formula in quantum mechanics is due to
the fact that the GPE is a nonlinear equation).  A WKB approximation
for the nonlinear Schr\"{o}dinger equation in an external potential,
uniformly valid in classically allowed and forbidden regions, was
recently proposed in Ref.~\cite{Japan}, but that result does not
produce the connection formula.  We develop an effectively 1D
treatment of the dynamics in cylindrical symmetry, which properly
accounts for the 3D aspects of the problem, see Sec.~\ref{S_Semicl}
below.  In the classically allowed region, the full 3D GPE is reduced
to a 1D counterpart, the solution to this 1D equation being subject to
a \emph{non-canonical} (non-quadratic) normalization condition [see
Eq.~(\ref{abnormal}) below], which is derived from the canonical
normalization for the 3D wave function.  As a physically relevant
application of the general method outlined above, in
Sec.~\ref{S_Momentum_Distribution} we calculate the quantum-mechanical
distribution of values of the longitudinal momentum in a cigar-shaped
BEC. This distribution also has a non-canonical form, in comparison
with ordinary quantum mechanics in 1D. In Sec.~\ref{S_Vorticity} we
generalize the approach to treat the case of a BEC with vorticity; in
this case, the effective 1D normalization condition takes on a still
more involved form [see Eq.~(\ref{exotic})].  We also develop a
generalization for the case when the cylindrical symmetry is broken by
an external potential, so that the effective equation is a 2D one, see
Sec.~\ref{S_Broken} below.  In that case, the 2D wave function is
subject to a different non-canonical normalization, see
Eq.~(\ref{abnormal2D}) below.  In Sec.~\ref{S_Numerical_Results} we
provide numerical examples that compare our method with full 3D
calculations for a BEC is static potentials and for dynamics of a BEC 
in the presence of a time-varying external potential.
 
We stress that our reduction of the 3D equation to its 1D (or 2D, in the 
broken-symmetry case) counterpart does not resort to the Gaussian 
approximation for the dependence of the wave function on the transverse 
coordinate(s) in the high-density region, where this approximation is not 
warranted. This is a principal difference from the approach developed in 
Refs.~\cite{Pedri,Salasnich}. 
 
\section{A Model with Cylindrical Symmetry} \label{S_Model} 
 
We begin by defining the kinds of systems of interest to us in the context 
of a cylindrically symmetric model problem. To this end, we consider a BEC 
in an array of optical traps, in the presence of the gravitational field and 
large-size magnetic trap in the form of a parabolic potential induced by the 
interaction of the magnetic moment of atoms with an external static magnetic 
field. The static magnetic-trap potential is  
\begin{equation} 
V_{M}(\mathbf{r})=\frac{1}{2}m\left[ \omega _{z}^{2}z^{2} +
\omega_{\perp }^{2}(x^{2}+y^{2})\right] , \label{trap}
\end{equation} 
where $m$ is the atomic mass.  The optical potential is produced by
light beams with identical linear polarizations, whose propagation
directions lie in one plane with the $z$ axis, forming angles $\theta
/2$ and $\pi -\theta/2$ with it, ($\theta =0$ corresponds to two beams
counter-propagating along the $z$ axis).  Interference between these
fields produces a standing-wave potential along the $z$ direction,
whose amplitude is proportional to the intensity of light.  It is
assumed that the intensity is initially zero and gradually increases
with time, hence the light-induced potential experienced by atoms in
the BEC is
\begin{equation} 
V_{L}(z,t)=V_{0}(t)\left[ 1+\cos (2kz)\right] /2,  \label{light} 
\end{equation} 
where $V_{0}(t)$ is the optical-potential's amplitude which varies in time, 
and $k\equiv \left( 2\pi /\lambda _{\mathrm{ph}}\right) \sin (\theta /2)$ is 
the wave vector of the optical lattices, $\lambda _{\mathrm{ph}}$ being the 
wavelength of light. It is often convenient to discuss the depth of the 
optical potential in units of the recoil energy $E_{R}=\left( 2\pi \hbar 
/\lambda _{\mathrm{ph}}\right) ^{2}/(2m)$, which is the kinetic energy 
gained by an atom when it absorbs a photon from the optical lattice. 
 
Finally, the description of the mean-field dynamics of the condensate is 
based on the 3D time-dependent GPE,  
\begin{equation} 
i\hbar \frac{\partial }{\partial t}\Psi (\mathbf{r},t)=\left[ \hat{p} 
^{2}/2m+V(\mathbf{r},t)+NU_{0}|\Psi |^{2}\right] \Psi \,,  \label{GP} 
\end{equation} 
where $V(\mathbf{r},t)=V_{M}(\mathbf{r})-mgz+V_{L}(z,t)$ is the full 
potential (the second term is the gravitational potential), $\hat{p}^{2}/2m$ 
is the kinetic-energy operator, $U_{0}=4\pi a_{0}\hbar ^{2}/m$ is the 
atom-atom interaction strength that is proportional to the $s$-wave 
scattering length $a_{0}$, and $N$ is the total number of atoms. Note that, 
according to Eq.~(\ref{trap}), the full potential can be written as  
\begin{equation} 
V(\mathbf{r},t)\equiv V_{z}(z,t)+V_{\perp }(\mathbf{r}_{\perp }),\,\,\mathrm{ 
\ with}\,\,V_{\perp }(\mathbf{r}_{\perp })\equiv \frac{1}{2}m\omega _{\perp 
}^{2}\left( x^{2}+y^{2}\right) .  \label{zperp} 
\end{equation} 
 
Equation~(\ref{GP}) can be rewritten, in terms of characteristic
diffraction and nonlinear time-scales $t_{DF}$ and $t_{NL}$, as
follows \cite{Tripp_2000,Tripp98}:
\begin{equation} 
\frac{\partial \Psi }{\partial t}=i\left[ \frac{r_{\mathrm{TF}}^{2}}{t_{ 
\mathrm{DF}}}\,\nabla ^{2}-V(\mathbf{r},t)/\hbar -\frac{1}{t_{\mathrm{NL}}} 
\, \frac{|\Psi |^{2}}{|\Psi _{m}|^{2}}\right] \Psi \ .  \label{GP_reduced} 
\end{equation} 
Here, the diffraction time $t_{DF}\equiv 2mr_{\mathrm{TF}}^{2}/\hbar
$, with $r_{\mathrm{TF}}=\sqrt{2\mu /(m\bar{\omega}^{2})}$ and
$\bar{\omega}=(\omega _{z}\omega _{\perp }^{2})^{1/3}$, and the
nonlinear time $t_{\mathrm{NL} }\equiv (G|\Psi _{m}|^{2}/\hbar
)^{-1}=(\mu /\hbar )^{-1}$ can be expressed in terms of the chemical
potential, where $|\Psi _{m}|$ is the maximum magnitude of $\Psi $
\cite{Tripp_2000}.  Another useful length scale is $r_{\mathrm{TF,z}}
= \sqrt{2\mu /(m\omega _{z}^{2})}$, i.e., the TF radius in the $z$
direction.  This is the size of a TF wave function in the $z$
direction for the harmonic potential $m\omega _{z}^{2}z^{2}/2$.
 
\section{The Semiclassical Approximation} \label{S_Semicl} 
 
We consider a semiclassical approach based on the Thomas-Fermi approximation 
for a 3D BEC wave function in a cylindrically symmetric potential. Our 
treatment is broken up into different approaches depending upon whether the 
atomic gas density is high (in the classically allowed regions of the 
coordinate space) or low (in classically forbidden regions). 
 
\subsection{Classically Allowed Region} 
 
In the classically allowed region [not too close to its boundaries so that 
the atomic gas density, and hence, the nonlinear term in Eq.~(\ref{GP}), 
remain sufficiently large], our description is based on the following  
\textit{ansatz} for the 3D wave function $\Psi (\mathbf{r},t)$,  
\begin{equation} 
\Psi (\mathbf{r},t)=\psi (z,t)\left( \frac{G-V_{\perp }(\mathbf{r}_{\perp 
})/|\psi (z,t)|^{2}}{G}\right) ^{1/2}  \label{ansatz} 
\end{equation} 
where $\psi (z,t)$ is a newly defined effective 1D wave function, and  
$G\equiv NU_{0}$ [cf.~Eq.~(\ref{GP})]. Note that the ansatz assumes a fairly 
simple relation between the squared 1D and 3D wave functions,  
\begin{equation} 
\left| \psi \right| ^{2}=V_{\perp }/G+\left| \Psi \right| ^{2}\,. 
\label{squared} 
\end{equation} 
Of course, the ansatz (\ref{ansatz}) makes sense in the region where  
\begin{equation} 
G|\psi (z,t)|^{2}-V_{\perp }(\mathbf{r}_{\perp })>0.  \label{necessary} 
\end{equation} 
 
Substituting the ansatz (\ref{ansatz}) into the 3D GPE and neglecting the 
transverse part of the kinetic-energy operator in the spirit of the TF 
approximation, we arrive at an effective 1D GPE,  
\begin{equation} 
i\hbar \frac{\partial \psi }{\partial t}=\left[ \frac{1}{2m}\hat{p} 
_{z}^{2}+V_{z}(z,t)+G|\psi |^{2}\right] \psi \,,  \label{1DGPE} 
\end{equation} 
upon neglecting terms proportional to $V_{\perp }$.  Strictly
speaking, to derive Eq.~(\ref{1DGPE}) we need a condition $V_{\perp
}\ll G|\Psi |^{2}$, in terms of the 3D wave function.  However, it
will be shown below that the 1D equation (\ref{1DGPE}) is a reasonable
approximate model even when the terms $V_{\perp }$ and $G|\Psi |^{2}$
are on the same order of magnitude.  The full TF approximation for
stationary solutions to Eq.~(\ref{1DGPE}) can be obtained, as
discussed in Sec.~\ref{FTFA} below, if one further neglects the
longitudinal kinetic-energy operator $\left( 1/2m\right)
\hat{p}_{z}^{2}$.  We stress that ansatz (\ref{ansatz}) is \emph{not}
an exact solution of the full 3D GPE, and it cannot describe radial
excitations of the BEC.
 
For the case of a harmonic transverse potential $V_{\perp }$ from
Eq.~(\ref{zperp}), the above condition (\ref{necessary}) yields
\begin{equation} 
r^{2}<r_{m}^{2}(z,t) \equiv \frac{2G}{m\omega_{\perp }^{2}}|\psi (z,t)|^{2}, 
\label{rmax} 
\end{equation} 
where $r$ is the radial coordinate in the plane $\left( x,y \right)$. The 
quantity $r_{m}$ introduced in Eq.~(\ref{rmax}) may be regarded as a 
definition of the radius of the cigar-shaped BEC at a given values of $z$ 
and $t$. The 3D wave function $\Psi (r,t)$ is subject to the ordinary 
normalization condition,  
\begin{equation} 
2\pi \int_{0}^{\infty }rdr\int_{-\infty }^{+\infty }dz\,|\Psi (r,z,t)|^{2}=1. 
\label{nirmul} 
\end{equation} 
Upon noting that the integration over $r$ in Eq.~(\ref{nirmul}) is confined 
to the region $r<r_{m}$ and substitution of the ansatz (\ref{ansatz}) into 
Eq.~(\ref{nirmul}), we obtain a result,  
\begin{eqnarray} 
2\pi \int_{-\infty }^{+\infty }dz\int_{0}^{r_{m}}rdr\,|\Psi (r,z,t)|^{2} &=&  
\frac{2\pi }{G}\int_{-\infty }^{+\infty }dz\int_{0}^{r_{m}}rdr\left[ G|\psi 
(z,t)|^{2}-m\omega _{\perp }^{2}r^{2}/2\right]  \nonumber \\ 
&\equiv &\frac{\pi m\omega _{\perp }^{2}}{G}\int_{-\infty }^{+\infty 
}dz\int_{0}^{r_{m}}rdr\left( r_{m}^{2}-r^{2}\right) =\frac{\pi G}{m\omega 
_{\perp }^{2}}\int_{-\infty }^{+\infty }|\psi (z,t)|^{4}dz=1.  \label{psi^4} 
\end{eqnarray} 
Thus, according to Eq.~(\ref{psi^4}), the usual normalization condition for 
the 3D wave function, Eq.~(\ref{nirmul}), generates the following \emph{\ 
non-canonical} normalization condition for the effective 1D wave function:  
\begin{equation} 
\int_{-\infty }^{+\infty }|\psi (z,t)|^{4}dz= \frac{m\omega _{\perp}^{2}}{ 
\pi G}\,.  \label{abnormal} 
\end{equation} 
We stress that this abnormal-looking condition is a direct result of
the standard full normalization condition (\ref{nirmul}) and the
ansatz (\ref{ansatz}) adopted for the 3D wave function.
 
We note that, as follows from Eq.~(\ref{ansatz}), $\psi (z,t)\equiv \Psi 
(x=0,y=0,z,t)$, i.e., the function $\psi $ is the particular value of the 
full wave function $\Psi $ on the axis $x=y=0$ (therefore, the functions  
$\psi $ and $\Psi $ are measured in the same units). Despite the on-axis 
identity between the functions $\Psi $ and $\psi $, the latter one does not 
have the interpretation as a probability amplitude for the distribution of 
atoms along the $z$ axis; instead, the probability for finding a particle in 
the region between $z$ and $z+dz$ (integrated in the transverse plane) is  
\begin{equation} 
P(z)\,dz=2\pi \int_{0}^{\infty }rdr\,|\Psi (r,z,t)|^{2}\,dz=\frac{\pi G}{ 
m\omega _{\perp }^{2}}|\psi (z,t)|^{4}dz\,,  \label{dz_abnormal} 
\end{equation} 
cf. Eqs.~(\ref{psi^4}) and (\ref{abnormal}). 
 
Recall that, in contrast to linear quantum mechanics, the normalization of 
the wave function $\psi $ is important and affects physical results in 
nonlinear theories of the GPE type. Indeed, the strength of the nonlinear 
mean-field term in Eq.~(\ref{1DGPE}) is determined by the maximum value of  
$|\psi |^{2}$ and is thus affected by the normalization. 
 
\subsection{Stationary and Slowly Varying Cases} 
 
A stationary solution to the 3D GPE with a time independent potential $V( 
\mathbf{r})$ can be approximated, in the classically allowed regions, by a 
stationary version of the ansatz (\ref{ansatz}), $\psi (r,z,t)=\phi 
(z)\,\exp \left[ -\left( i/\hbar \right) \mu t\right] $, where the function  
$\phi (z)$ satisfies the equation  
\begin{equation} 
-\frac{\hbar ^{2}}{2m}\frac{d^{2}\phi }{dz^{2}}+\left[ V_{z}(z)-\mu \right] 
\phi +G\phi ^{3}=0\,  \label{phi} 
\end{equation} 
following from Eq. (\ref{1DGPE}). 
 
For problems with a slow time variation, we can consider an instantaneous 
eigenstate of the nonlinear time-dependent GPE equation. Adiabatically 
varying potentials $V_{z}(z,t)$ can be treated by calculating the 
instantaneous chemical potential and quasi-stationary wave function $\phi 
(z;t)$ in the instantaneous external potential, and then forming the full 
time-dependent solution as  
\begin{equation} 
\psi (r,z,t)=\phi (z;t)\,\exp \left[ -\left( i/\hbar \right) 
\int_{0}^{t}dt^{\prime }\mu (t^{\prime })\right] .  \label{quasi_phi} 
\end{equation} 
Strictly speaking, the non-canonical normalization condition (\ref{abnormal}), 
unlike the canonical (quadratic) one, is not compatible with the full 
time-dependent effective 1D GPE (\ref{1DGPE}). However, there is no problem 
with the compatibility in the case of the adiabatically slow evolution. 
 
\subsection{Full Thomas-Fermi Approximation in the Classically Allowed
Region} \label{FTFA}
 
The full 3D TF approximation can be recovered if we apply the TF 
approximation directly to the 1D equation (\ref{1DGPE}), neglecting the 
kinetic-energy operator in it, so that the solution will be  
\begin{equation} 
\psi (z,t)=\sqrt{\frac{\mu -V_{z}(z,t)}{G}}\,\exp \left[ -\frac{i}{\hbar } 
\int_{0}^{t}dt^{\prime }\mu (t^{\prime })\right] \,.  \label{fullTF} 
\end{equation} 
According to Eq. (\ref{ansatz}), this yields the full 3D TF wave function,  
\begin{equation} 
\Psi (\mathbf{r},t)=\left( \frac{\mu -V_{\perp }(\mathbf{r})_{\perp 
}-V_{z}(z,t)}{G}\right) ^{1/2}\,\exp \left[ -\frac{i}{\hbar } 
\int_{0}^{t}dt^{\prime }\mu (t^{\prime })\right] \,.  \label{full_TF_PSI} 
\end{equation} 
Substituting Eq.~(\ref{fullTF}) into the non-canonical normalization 
condition (\ref{abnormal}) yields  
\begin{equation} 
\int_{-\infty }^{+\infty }\left[ \mu (t)-V_{z}(z,t)\right] ^{2}dz=Gm\omega 
_{\perp }^{2}/\pi \,,  \label{mu-condition} 
\end{equation} 
where the region of integration over $z$ is restricted by the
condition $\mu (t)-V_{z}(z,t)>0$.  Hence, the normalization condition
(\ref{mu-condition}) determines the chemical potential $\mu (t)$. 
Note that the condition (\ref{mu-condition}) is equivalent to the
usual form of the condition which determines the chemical potential in
the framework of the TF approximation applied to the full 3D equation
(\ref{GP}):
\begin{equation} 
\frac{2\pi }{G}\int_{0}^{\infty }rdr\int_{-\infty }^{+\infty }dz\,\left[ \mu 
(t)-V(\mathbf{r},t)\right] =1,  \label{nirmul-mu} 
\end{equation} 
where the integration is performed over the region in which $\mu (t)-V( 
\mathbf{r},t)>0$. Performing the algebra, we arrive at the usual expression 
for the chemical potential in the static harmonic 3D potential (without an 
optical component):  
\[ 
\mu =\frac{1}{2}\left[ 15G/(4\pi )\right] ^{2/5}(m\bar{\omega}^{2})^{3/5}\,. 
\] 
Finally, for any potential $V_{z}(z)$ (and the harmonic potential $V_{\perp } 
$), the effective probability density defined by Eq. (\ref{dz_abnormal}) 
takes the following form in the TF approximation:  
\begin{equation} 
P_{\mathrm{TF}}(z)=\frac{\pi }{m\omega _{\perp }^{2}G}|\mu -V_{z}(z)|^{2},\, 
\label{Prob_TF} 
\end{equation} 
in the region where$\,\mu -V_{z}(z)>0$; otherwise, $P_{\mathrm{TF}}(z)=0$. 
 
\subsection{Classically Forbidden Regions} 
 
In the classically forbidden regions, the density of atoms is small, 
therefore the nonlinear term in the GPE may be dropped, so that it becomes 
tantamount to the ordinary quantum-mechanical Schr\"{o}dinger equation, 
hence we adopt the following product ansatz for $\Psi (\mathbf{r},t)$:  
\begin{equation} 
\Psi (\mathbf{r},t)=\psi (z,t)\,\exp (-\mathbf{r}_{\perp }^{2}/2R_{\perp 
}^{2}-i\omega _{\perp }t/2)\,,  \label{ansatz'} 
\end{equation} 
where the transverse squared radius is $R_{\perp }^{2}=\hbar /m\omega 
_{\perp }$. We stress that the Gaussian approximation for the transverse 
part of the ansatz (\ref{ansatz'}) is appropriate, unlike in the classically 
allowed region, as the equation is effectively linear in the present case. 
Upon substituting Eq.~(\ref{ansatz'}) into the linearized GPE, it is 
straightforward to obtain an effective 1D linear Schr\"{o}dinger equation,  
\begin{equation} 
i\hbar \frac{\partial \psi (z,t)}{\partial t}=\left[ \frac{1}{2m}\hat{p}{\ 
_{z}^{2}}+V_{z}(z,t)\right] \psi \,.  \label{1DSchE} 
\end{equation} 
Quite naturally, Eq.~(\ref{1DSchE}) is equivalent to Eq.~(\ref{1DGPE}) in 
the classically forbidden region, as in this region the nonlinear term in 
Eq.~(\ref{1DGPE}) is negligible. 
 
\section{Matching Conditions} \label{S_Matching_Conditions} 
 
Equations~(\ref{1DGPE}) or (\ref{GP_reduced}) can be rewritten in the 
following form by rescaling $t$, $z$ and $\psi $:  
\begin{equation} 
iu_{t}+\frac{1}{2}u_{zz}-U(z,t)u-\gamma \left| u\right|^{2}u=0\,, 
\label{1GP} 
\end{equation} 
where $\gamma >0$ is a properly normalized nonlinear coupling strength. A 
solution to Eq.~(\ref{1GP}) is sought for as  
\begin{equation} 
u(z,t)=v(z;t)\exp \left( -i\int_{0}^{t}dt^{\prime }\mu (t^{\prime })\right) , 
\label{v} 
\end{equation} 
with a real chemical potential $\mu (t)$ and a real function $v(z;t)$, cf. 
Eq. (\ref{quasi_phi}). Here, the time dependence of $U(z,t)$ is presumed to 
be slow enough, and $v(z;t)$ satisfies a quasi-stationary equation,  
\begin{equation} 
\left[ \mu (t)-U(z,t)\right] v+\frac{1}{2}\frac{d^{2}v}{dz^{2}}-\gamma 
v^{3}=0.  \label{stationary} 
\end{equation} 
 
As in the ordinary semiclassical form of quantum mechanics, it is
necessary to match the approximations for the wave function across the
classical turning point, which separates the classically allowed and
forbidden regions in the 1D space.  As well as in linear quantum
mechanics, the wave function in the latter region will be taken in the
WKB approximation, see Eq.~(\ref{under}) below.  However, a crucial
difference from the standard theory is that the wave function in the
classically allowed area is taken not in the corresponding version of
the WKB approximation, but rather in the TF form.  This, of course,
drastically changes the matching problem (see also Ref.~\cite{Japan}).
 
Deeply under the barrier, i.e., for large positive values of the potential  
$U(z)$, the density of particles is small, hence, as it was already mentioned 
above, the nonlinear term in Eq.~(\ref{stationary}) may be dropped, and a 
solution may be presented in the standard semi-classical (WKB) 
approximation,  
\begin{equation} 
v(z)=\frac{C}{\left[ 2\left( U(z)-\mu \right) \right]^{1/4}}\exp \left[ 
-\int_{z_{0}}^{z}\sqrt{2\left( U(z^{\prime })-\mu \right) }\,dz^{\prime } 
\right] ,  \label{under} 
\end{equation} 
where, for the definiteness, we choose $z_{0}$ as the classical turning 
point, at which $U(z_{0})=\mu $, and $C$ is (for the time being) an 
arbitrary real constant (because $C$ is arbitrary, $z_{0}$ may indeed be 
chosen arbitrarily). It is also assumed that the classically forbidden 
region is located at $z>z_{0}$, i.e., to the right of the turning point. As 
usual, the WKB approximation (\ref{under}) is not valid too close to the 
turning point. 
 
On the other hand, the solution in the classically allowed region  
($z<z_{0}$), not too close to the turning point,  
is taken in the usual TF 
approximation as described above, cf. Eq.~(\ref{fullTF}):  
\begin{equation} 
v_{\mathrm{TF}}=\sqrt{\frac{\mu -U(z)}{\gamma }}.  \label{TF} 
\end{equation} 
Note that, unlike the WKB solution (\ref{under}), the TF approximate 
solution (\ref{TF}) does not contain any arbitrary constant. 
 
Now we need to match the two approximations (\ref{under}) and
(\ref{TF}) across the turning point, in a vicinity of which both
approximations are not applicable, the eventual objective being to
find the constant $C$ in Eq.~(\ref{under}).  Following the usual
quantum-mechanical approach, one can cast the matching problem into a
standard form, expanding the potential in a vicinity of the turning
point,
\begin{equation} 
\mu -U(z)\approx F_{0}\left( z-z_{0}\right) ,  \label{F} 
\end{equation} 
where $F_{0}$ is the value of the potential force at the turning point
(in the present case, $F_{0}<0$).  The accordingly modified version of
Eq.~(\ref{stationary}) is
\[ 
\frac{d^{2}v}{dz^{2}}+F_{0}\left( z-z_{0}\right) v-2\gamma v^{3}=0,  
\] 
which is transformed into a normalized form,  
\begin{equation} 
\frac{d^{2}w}{d\xi^{2}}=\xi w+2w^{3},  \label{Painleve'} 
\end{equation} 
by means of rescalings  
\begin{equation} 
\xi \equiv \left( 2\left| F_{0}\right| \right)^{1/3}\left( z-z_{0}\right) 
,\,\,v \equiv \frac{\left( 2\left| F_{0}\right| \right)^{1/3}}{\sqrt{\gamma } 
}w.  \label{rescaling} 
\end{equation} 
 
Equation (\ref{rescaling}) is a particular case of a classical equation 
known as the Painlev\'{e} transcendental of the second type. The full form 
of this equation (a standard notation for which is \textrm{P}$_{\mathrm{II}}$) 
is  
\[ 
\frac{d^{2}w}{d\xi ^{2}}=\xi w+2w^{3}+\alpha , 
\] 
cf.~Eq.~(\ref{Painleve'}), where $\alpha $ is an arbitrary real parameter; 
in the present case, $\alpha \equiv 0$. The use of the expansion (\ref{F}) 
and simplified equation (\ref{Painleve'}) for matching different asymptotic 
solutions of an equation equivalent to Eq.~(\ref{stationary}) was proposed, 
in a context different from BEC, in Ref.~\cite{Haberman} (however, the 
asymptotic form of the solution in the classically allowed region, for which 
the analysis was done in Ref.~\cite{Haberman}, was different from Eq.~(\ref 
{TF}): it corresponded to a nonlinear wave function oscillating in space, 
rather than to the TF case). Here, it is necessary to find a solution to 
Eq.~(\ref{Painleve'}) with the property that it takes the asymptotic forms  
\begin{equation} 
w\approx \sqrt{-\xi /2}\,\,\mathrm{and}\,\,\,\,w\approx \frac{\widetilde{C}}{ 
\xi ^{1/4}}\exp \left( -\frac{2}{3}\xi ^{3/2}\right)   \label{asympt} 
\end{equation} 
at $\xi \rightarrow -\infty $ and $\xi \rightarrow +\infty $, respectively, 
in accordance with Eqs.~(\ref{TF}) and (\ref{under}). The actual problem is 
to find a value of the universal numerical constant $\widetilde{C}$ in  
Eqs.~(\ref{asympt}) from matching the solution to its  
uniquely defined asymptotic 
form at $\xi \rightarrow -\infty $. 
 
As mentioned above, this matching problem is different from its counterpart 
in ordinary (linear) quantum mechanics. An \emph{exact solution} to this 
problem is available in the mathematical literature (see Ref. \cite{AS} and 
references to original works therein). The final result of the analysis is  
\begin{equation} 
\widetilde{C}=\frac{1}{2\sqrt{\pi }}\approx 0.\,\allowbreak 282\,. 
\label{exact} 
\end{equation} 
Using this exact result, and undoing the rescalings (\ref{rescaling}), we 
obtain the value of the constant $C$ in the WKB solution (\ref{under}):  
\begin{equation} 
C=\widetilde{C}\sqrt{\frac{2\left| F_{0}\right| }{\gamma }}\equiv \sqrt{ 
\frac{\left| F_{0}\right| }{2\pi \gamma }}.  \label{C} 
\end{equation} 
With the relation (\ref{C}), the WKB expression (\ref{under}) for the wave 
function under the barrier is completely defined. Of course, the result 
presented in the form (\ref{C}) also applies when the classically forbidden 
region is located to the left (rather than to the right) of the classical 
turning point. 
 
This result for the matching problem applies as well to the case of the 
adiabatically slow variation of the parameters, e.g., when $F_{0}$ slowly 
varies as a function of time (or when $\gamma $ changes with time due to 
variation of the trap potential with time). The result does not apply to the 
case of a vertical potential wall (when, formally, $F_{0}=\infty $). 
However, in this case, the solution is almost trivial. Indeed, assume that 
at the point $z=0$ there is a jump of the potential from a large negative 
value $U_{-}$ inside the well ($z<0$) to a large positive value $U_{+}$ in 
the classically forbidden region ($z>0$). Then, in the allowed region, the 
TF solution in the form $v=\sqrt{(\mu -U_{-})/\gamma }$ is valid everywhere 
up to the turning point, the exact solution in the forbidden region is  
\[ 
v=C\exp \left( -\sqrt{2(U_{+}-\mu )}z\right) ,  
\] 
and the constant $C$ is immediately found from the continuity condition, $C= 
\sqrt{(\mu -U_{-})/\gamma }$. 
 
\section{Momentum Distribution} \label{S_Momentum_Distribution} 
 
The approach developed above can be naturally applied to calculate the 
distribution of values of the longitudinal momentum $p$ in a given quantum 
state $|\Psi \rangle $, which can be measured in a direct experiment. The 
distribution is determined by the scalar product  
\begin{equation} 
\mathcal{P}(p)=\left| \langle p|\Psi \rangle \right| ^{2}\,,  \label{rho} 
\end{equation} 
where, up to a normalization factor,  
\begin{equation} 
\langle p|=\exp \left( \frac{-ipz}{\hbar }\right) \,  \label{left} 
\end{equation} 
is the conjugate eigenfunction of the momentum operator. Thus, according to 
Eq. (\ref{rho}), to determine the momentum distribution in the $z$ direction 
we need to calculate nothing else but the 1D Fourier transform of the 3D 
wave function (\ref{ansatz}), additionally integrated in the transverse 
direction [the scalar product in Eq.~(\ref{rho}) assumes, of course, the 
full 3D integration]:  
\begin{equation} 
\langle p|\Psi \rangle =2\pi \int_{-\infty }^{+\infty }dz\exp \left( \frac{ 
-ipz}{\hbar }\right) \,\,\int_{0}^{\infty }\Psi (r_{\perp },z,t)\,r_{\perp 
}dr_{\perp }\,,  \label{Fourier} 
\end{equation} 
where the multiplier $2\pi $ is generated by the angular integration in the 
transverse plane. 
 
Substituting the expression (\ref{ansatz}) for $\Psi $ in the classically 
allowed region into (\ref{Fourier}), the integration over $r_{\perp }$ is 
confined to the interval $0<r_{\perp }<r_{m}$, where $r_{m}^{2}(z,t)$ is the 
same as defined above by Eq. (\ref{rmax}). Taking into account the form of 
the transverse potential (\ref{zperp}), we arrive at an expression  
\begin{eqnarray} 
\langle p|\Psi \rangle &\approx &2\pi \int_{-\infty }^{+\infty }dz\exp 
\left( \frac{-ipz}{\hbar }\right) \,\,\psi (z,t)\int_{0}^{r_{m}}\,\left(  
\frac{G-V_{\perp }(\mathbf{r}_{\perp })/|\psi (z,t)|^{2}}{G}\right) 
^{1/2}r_{\perp }dr_{\perp }  \nonumber \\ 
&=&2\pi \int_{-\infty }^{+\infty }r_{m}^{2}\,dz\exp \left( \frac{-ipz}{\hbar  
}\right) \psi (z,t)\int_{0}^{1}\sqrt{1-\rho ^{2}}\rho d\rho \,,  \label{int} 
\end{eqnarray} 
where $\rho \equiv r_{\perp }/r_{m}$. Further, using an elementary formula  
$\int_{0}^{1}\sqrt{1-\rho ^{2}}\rho d\rho =\allowbreak 1/3$ and the 
expression (\ref{rmax}) for $r_{m}$, we obtain from Eq. (\ref{int}),  
\begin{equation} 
\langle p|\Psi \rangle =\frac{4\pi }{3}\frac{G}{m\omega _{\perp }^{2}} 
\int_{-\infty }^{+\infty }dz\exp \left( \frac{-ipz}{\hbar }\right) \psi 
(z,t)|\psi (z,t)|^{2}\,,  \label{pPsi} 
\end{equation} 
which is the final result: Eq. (\ref{pPsi}) tells one that the amplitude  
$\langle p|\Psi \rangle $ of the probability distribution, that determines 
the probability as per Eq.~(\ref{rho}), is  
\begin{equation} 
\langle p|\Psi \rangle =\frac{4\pi (2\pi )^{1/2}}{3}\frac{G}{m\omega _{\perp 
}^{2}}\mathcal{F}\{\psi \left| \psi \right| ^{2}\}(p/\hbar )\,, 
\label{final} 
\end{equation} 
where $\mathcal{F}$ is the symbol of the Fourier transform,  
\begin{equation} 
\mathcal{F}\{f\}(\omega )\equiv \frac{1}{(2\pi )^{1/2}}\int_{-\infty 
}^{+\infty }\exp \left( -i\omega t\right) \,f(t)dt\,.  \label{f} 
\end{equation} 
In the classically forbidden region, the usual quantum mechanical 
expressions for the momentum distribution apply. 
 
Expectation values of operators involving only the longitudinal momentum can 
be computed as follows, if we neglect a contribution from the classically 
forbidden regions. Suppose we have an operator $\mathcal{O}(p)$, such as, 
for instance, $p$ itself or the kinetic energy, $p^{2}/\left( 2m\right) $. 
Then, the expectation value of the operator $\mathcal{O}(p)$ is given by  
\begin{equation} 
\langle \Psi |\mathcal{O}(p)|\Psi \rangle =\int_{-\infty }^{+\infty }dp 
\mathcal{O}(p)\left| \langle p|\Psi \rangle \right| ^{2}\,.  \label{expect} 
\end{equation} 
The expression~(\ref{final}) should be then substituted for $\langle p|\Psi 
\rangle $ in Eq.~(\ref{expect}). 
 
\section{The Condensate with Vorticity} \label{S_Vorticity} 
 
The above consideration can be generalized for a case when the condensate 
inside the cylindrically symmetric region is given vorticity, so that the 
wave function has the form  
\begin{equation} 
\Psi (r,z,t)=\exp \left( il\theta \right) \,\Phi (r,z,t),  \label{vortex} 
\end{equation} 
where $\theta $ is the angular coordinate in the transverse plane, and $l$ 
is the vorticity quantum number. The ansatz (\ref{ansatz}) is then replaced 
by  
\begin{equation} 
\Psi (r,z,t)=\exp \left( il\theta \right) \,G^{-1/2}\sqrt{G|\psi 
_{l}(z,t)|^{2}-\left( m\omega _{\perp }^{2}r^{2}/2+\frac{\hbar ^{2}l^{2}} 
{2mr^{2}}\right) }\,.  \label{generalizedAnsatz} 
\end{equation} 
The corresponding normalization condition replacing Eq.~(\ref{nirmul}) takes 
the form  
\begin{equation} 
2\pi \int_{-\infty }^{+\infty }dz\int_{r_{\min }}^{r_{m}}rdr\,|\Psi 
(r,z,t)|^{2}=1,  \label{minmax} 
\end{equation} 
where now,  
\begin{eqnarray} 
r_{m}^{2} &=&\frac{1}{m\omega _{\perp }^{2}}\left[ G\left| \psi _{l}\right| 
^{2}+\sqrt{G^{2}\left| \psi _{l}\right| ^{4}-\left( l\hbar \omega _{\perp 
}\right) ^{2}}\right] ,  \label{max} \\ 
r_{\min }^{2} &=&\frac{1}{m\omega _{\perp }^{2}}\left[ G\left| \psi 
_{l}\right| ^{2}-\sqrt{G^{2}\left| \psi _{l}\right| ^{4}-\left( l\hbar 
\omega _{\perp }\right) ^{2}}\right] .  \label{min} 
\end{eqnarray} 
The corresponding 1D GPE equation for $\psi _{l}(z,t)$ is  
\[ 
i\hbar \frac{\partial \psi _{l}}{\partial t}=\left[ \frac{1}{2m}\hat{p} 
_{z}^{2}+V_{z}(z,t)+G|\psi _{l}|^{2}\right] \psi _{l}\,. 
\] 
Finally, the substitution of the expressions (\ref{generalizedAnsatz})
and (\ref{max}), (\ref{min}) into Eq.~(\ref{minmax}) yields, after a
straightforward calculation of the integral over the radial variable
$r$, an effectively 1D normalization condition in a complicated form,
which is a generalization of the above non-canonical normalization
condition (\ref{abnormal}) corresponding to $l=0$:
\begin{eqnarray} 
&&\int_{-\infty }^{+\infty }dz\left[ \frac{G\left| \psi _{l}(z)\right| ^{2} 
\sqrt{G^{2}\left| \psi _{l}(z)\right| ^{4}-\left( l\hbar \omega _{\perp 
}\right) ^{2}}}{m^{2}\omega _{\perp }^{4}}\right.   \nonumber \\ 
&&\left. -\left( \frac{\hbar l}{m\omega _{\perp }}\right) ^{2}\ln \left(  
\frac{2G^{2}\left| \psi _{l}(z)\right| ^{4}}{\left( l\hbar \omega _{\perp 
}\right) ^{2}}-1+\frac{2G\left| \psi _{l}(z)\right| ^{2}}{l\hbar \omega 
_{\perp }}\sqrt{\frac{G^{2}\left| \psi _{l}(z)\right| ^{4}}{\left( l\hbar 
\omega _{\perp }\right) ^{2}}-1}\right) \right] =\frac{2G}{\pi m\omega 
_{\perp }^{2}}\,.  \label{exotic} 
\end{eqnarray} 
Note that, with $l=0$, this result indeed reduces to
Eq.~(\ref{abnormal}).  A consequence of Eq.~(\ref{exotic}) is that
$\left| \psi _{l}(z)\right| ^{2}$ cannot take values smaller than
$\left( \left| \psi _{l}(z)\right| ^{2}\right) _{\min }=\left( l\hbar
\omega _{\perp }\right) /G$.  Furthermore, from
Eq.~(\ref{generalizedAnsatz}) it is clear that the condition $G|\psi
_{l}(z,t)|^{2}-\left[ m\omega _{\perp }^{2}r^{2}/2+\left( \hbar
l/r\right) ^{2}/\left( 2m\right) \right] >0$ must be satisfied for the
ansatz (\ref{generalizedAnsatz}) to be valid.
 
\section{The Broken-Symmetry Condensate} \label{S_Broken} 
 
As mentioned in the introduction, the approach developed above can be 
extended to the case when the cylindrical symmetry about the $z$ axis is 
broken by a non-axisymmetric potential $V_{xy}(x,y)$. In this case, 
essentially the same ansatz as given by Eq.~(\ref{ansatz}) may be employed 
if the potential is harmonic in $z$. Accordingly, we take the transverse 
potential as $V_{\perp }(z)=\left( 1/2\right) m\omega _{\perp }^{2}z^{2}$, 
cf.~Eq.~(\ref{zperp}), and an ansatz in the form  
\begin{equation} 
\Psi (\mathbf{r},t)=\psi (x,y,t)\left( \frac{G-m\omega _{\perp 
}^{2}z^{2}/2|\psi (x,y,t)|^{2}}{G}\right) ^{1/2}\,.  \label{2D} 
\end{equation} 
Note that the interpretation of the function $\psi (x,y,t)$ is similar to 
that of the function $\psi (z,t)$ in the ansatz (\ref{ansatz}): it coincides 
with the full 3D wave function $\Psi (x,y,z,t)$ at $z=0$. Substitution of 
Eq.~(\ref{2D}) into the 3D normalization condition (\ref{nirmul}) and 
straightforward integration in the $z$--direction yields a result  
\begin{equation} 
\int_{-\infty }^{+\infty }\int_{-\infty }^{+\infty }\left| \psi (x,y)\right| 
^{3}dx\,dy=\frac{3}{4}\sqrt{\frac{m\omega _{\perp }^{2}}{2G}}, 
\label{abnormal2D} 
\end{equation} 
cf.~Eq.~(\ref{abnormal}).  This is another example of the
non-canonical normalization condition which is generated by the
reduction in the effective space dimension.  As for the effective 2D
GPE generated by the ansatz (\ref{2D}), it has the usual form,
\begin{equation} 
i\hbar \frac{\partial \psi (x,y,t)}{\partial t}=\left[ \frac{1}{2m}\left(  
\hat{p}_{x}^{2}+\hat{p}_{y}^{2}\right) +V_{xy}(x,y,t)\right] \psi \,, 
\label{2DGPE} 
\end{equation} 
cf.~Eq.~(\ref{1DGPE}) [a formal condition under which Eq.~(\ref{2DGPE}) can 
be derived from the 3D equation (\ref{GP}) is $m\omega _{\perp }^{2}z^{2}\ll 
|\psi (x,y,t)|^{2}$]. 
 
We do not consider the problem of matching the wave function in the 
classically allowed and forbidden regions in the framework of the 2D 
equation (\ref{2DGPE}), as the WKB approximation for the classically 
forbidden region is itself problematic in the 2D case. In fact, this 
approximation was only elaborated for the 2D motion in an axially symmetric 
field \cite{LLQM}, which does not correspond to the situation of interest in 
the present context [the axial symmetry in the $\left( x,y\right)$ plane 
will be destroyed by the optical lattice]. 
 
\section{Numerical Results} \label{S_Numerical_Results} 
 
For the numerical solution of Eqs.~(\ref{GP}) and (\ref{1DGPE}), we used the 
standard split-step operator method \cite{Fleck_Feit}. The computational 
grids had $65536$ and $32\times 32\times 2048$ points for the 1D and 3D 
cases, respectively, with spatial steps $\Delta h_{z}\approx 0.0002\,r_{ 
\mathrm{TF},z}$ for the 1D geometry, and $\Delta h_{x}=\Delta h_{y}\approx 
0.13\,r_{\mathrm{TF}}$ and $\Delta h_{z}\approx 0.005\,r_{\mathrm{TF}}$ for 
the 3D case. We used two different methods for finding stationary solutions 
to GPE~(\ref{phi}), and to its 3D counterpart: the first technique was an 
imaginary-time version of the split-step operator method \cite{Tripp_2000}, 
and the second is the standard finite-difference method used to solve a 
two-point boundary value problem \cite{Press}. It is important to note that, 
by treating Eq.~(\ref{phi}) as a two-point boundary problem, the initial 
conditions must include the eigenvalue $\mu $. This will, generally, give an 
unphysical solution for $\phi $ since the normalization condition will not 
be met. By gradually changing the value of $\mu $, and redetermining the 
wave function of the nonlinear GPE using the finite-difference method, we 
can follow the surface of solutions to Eq.~(\ref{phi}) until a physical 
solution is obtained. 
 
Numerically, we find the imaginary time split-step relaxation method 
painfully difficult to converge in 3D for the cigar-shaped geometry that we 
used in the calculations, both with and without the optical potential. The 
finite-difference two-point boundary value method appears to be more 
efficient. For dynamical simulations, however, we used only the split-step 
method. 
 
We consider a $^{87}$Rb condensate with $10^{6}$ atoms in a harmonic 
potential with trap frequencies $\omega _{z}=100$ Hz and $\omega _{x}=\omega 
_{y}=20$ Hz, and s-wave scattering length $a_{0}=5.017$ nm. We first 
consider the case when only the harmonic potential is present; the action of 
an optical potential on the trapped BEC will be considered below. For the 
parameters used in the calculations, the TF radius in the $z$ direction is  
$r_{\mathrm{TF},z}=36.35$ $\mu $m, and the chemical potential is $\mu 
=1.507\times 10^{-30}$ J. 
 
\begin{figure}[!htb] 
\centerline{ 
\includegraphics[width=3in,angle=270,keepaspectratio]{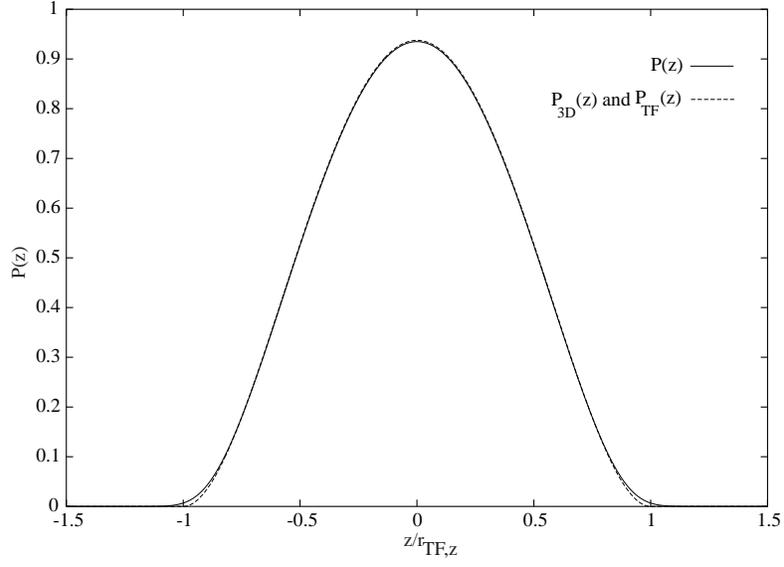}}  
\caption{The probability $P(z)$ for the 3D harmonic potential versus
$z/r_{ \mathrm{\ TF},z}$, as found numerically using
Eq.~(\ref{GP_reduced}), and $P_{ \mathrm{TF}}(z)$ versus
$z/r_{\mathrm{TF},z}$, as calculated using Eq.~(\ref {Prob_TF}).  The
probability $P_{3D}(z)$ found from the stationary version of the full
3D equation (see the text) cannot be discerned, as it lies on top of
$P_{\mathrm{TF}}(z)$.}
\label{fig1} 
\end{figure}

Figure~\ref{fig1} shows the probability $P(z)$ versus $z$, as calculated 
using Eq.~(\ref{GP_reduced}). Also shown is the probability $P_{\mathrm{TF} 
}(z)$ as obtained from the full TF approximation based on Eq.~(\ref{Prob_TF} 
), and a probability $P_{3D}(z)$ found from the numerical solution of the 
full 3D GPE. The curve obtained from the 3D GPE lies on top of the TF curve, 
being nearly undiscernible from it (it is no surprise that TF is a good 
approximation for $10^{6}$ Rb atoms in a harmonic trap). Clearly, the 
comparison of the $P(z)$ found by means of the approximation developed above 
with the TF and full 3D results is excellent. A slight deterioration occurs 
in the region where the density is low. 
 
\begin{figure}[!htb] 
\centerline{ 
\includegraphics[width=3in,angle=270,keepaspectratio]{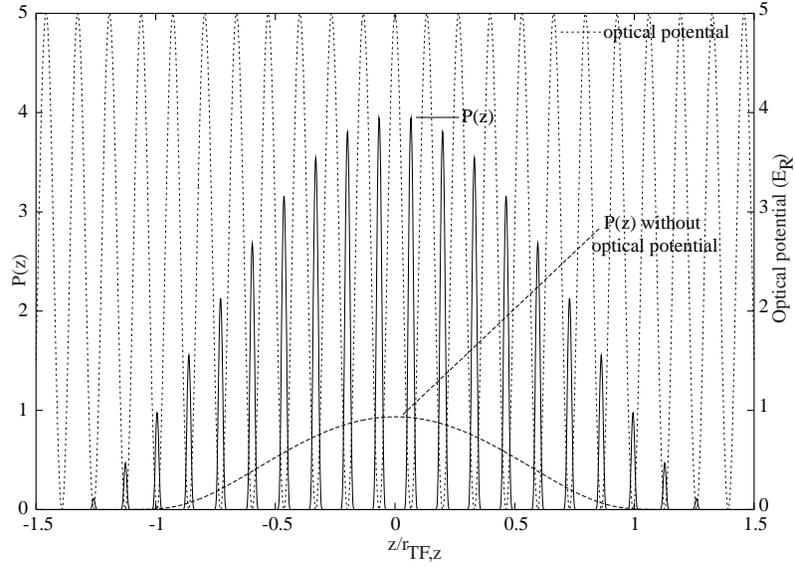}}  
\caption{The probability $P(z)$ for the combination of the 3D harmonic and 
optical potentials vs $z/r_{\mathrm{TF},z}$, as found numerically using 
Eq.~(\ref{GP_reduced}). For comparison, the probability $P(z)$ without the 
optical potential, and the optical potential itself are also shown.} 
\label{fig2} 
\end{figure}

Figure~\ref{fig2} shows the probability distribution for the same
harmonic potential as in Fig.~\ref{fig1}, but with an added
optical-lattice potential as given by Eq.~(\ref{light}), with the
wavelength $\lambda _{\mathrm{ph} }=840$ nm, relative angle $\theta
=10$ degrees between the two light beams, and the constant amplitude
$V_{0}=5\,E_{R}$.  For this wavelength, the recoil energy $E_{R}=\hbar
^{2}(2\pi /\lambda )^{2}/(2m)=2.15\times 10^{-30}$ J, so the
optical-potential's strength is about $1.4$ times the value of the
chemical potential in the absence of the optical lattice.  In the
presence of the optical lattice, the chemical potential is calculated
to be $3.15\times 10^{-30}$ J (slightly more than twice the chemical
potential without the optical lattice).  Also shown in Fig.~\ref{fig2}
are the optical potential $V_{L}(z)$, and $P(z)$ determined without
the optical lattice (as in Fig.~\ref{fig1}).  As is seen from the
figure, the optical potential squeezes the atomic density out to
larger $z$; it also squeezes it out to larger $x$ and $y$, see below.
 
Figure~\ref{fig3} shows the comparison of $P(z)$ with
$P_{\mathrm{TF}}(z)$ and $P_{3D}(z)$, where $P_{\mathrm{TF}}(z)$ is
calculated using Eq.~(\ref{Prob_TF}), and with $P_{3D}(z)$ as obtained
by solving the stationary 3D GPE. The TF result still provides an
adequate description, despite the fact that the kinetic energy is more
important in this case than in the case without the optical potential. 
Hence we conclude that the kinetic energy remains much less
significant than the potential and the mean-field energy in this case. 
Our method produces results closer to the 3D result than the TF. The
kinetic energy does broaden the wave function in each optical well,
hence peaks of the wave functions are lower than in the TF
approximation, as evident in Fig.~\ref{fig3}.  Note, however, that our
method, being based on a TF-type approximation in the transverse
dimension, restricts the diffusion of the wave function due to the
kinetic energy to the $z$ direction (but the wave function is
definitely squeezed into the transverse dimension due to the optical
potential and the mean-field -- see next paragraph).  The inset in the
figure is a blowup of the region near $z/r_{\mathrm{TF},z}=1$.
 
Figure~\ref{fig4} shows $|\Psi (\mathbf{r})|^{2}$ versus $r$ and $z$. 
A striking aspect of this figure is the extent to which the wave
function is squeezed out to larger $r$ in the presence of the optical
potential.  Without this potential, the size of the wave function in
$r$ is $r_{\mathrm{TF},x}$ ($=r_{\mathrm{TF},y}$), but now it is
squeezed out to about $5\,r_{\mathrm{TF} ,x}$.  The size of the wave
function in the radial direction depends upon $z$, as does the extent
of the squeezing in $r$.  The distribution of $|\Psi
(\mathbf{r})|^{2}$ versus $r$ and $z$, as produced by the stationary
3D GPE, is similar to that obtained by our method.
 
Lastly, in Fig.~\ref{fig5} we show results of a dynamical calculation
in which we varied the optical potential as a function of time, so
that the peak strength depended on time as $V_{0}(t)=5E_{R}\times \exp
[-\left( (t-t_{F})/\sigma \right) ^{2}]$, with $t_{F}=1.12$ ms and
$\sigma =t_{F}/2=0.56$ ms.  The calculated probability distribution
$P(z,t)$ is shown at four different times, viz., at $t=0$ before the
optical potential is ramped up, at $t=t_{F}/5$ when the optical
potential is still rather small, at $t=t_{F}/2$ when the optical
potential is somewhat less than 0.4 times its final value of $5E_{R}$,
and at $t=t_{F}$, when $V_{0}(t_{F})=5E_{R}$.  The dynamics are
initially adiabatic, but, clearly, by the final time, $t_{F}=1.12$ ms,
the dynamics cease to be adiabatic (compare the result with the
probability distribution shown in Fig.~\ref{fig2}).  The dynamics of
the probability $P(z,t)$ versus $z$, calculated using the 3D GPE, are
similar to that shown in Fig.~\ref{fig5}.
 
\section{Summary and Conclusions} \label{S_Summary} 
 
We have proposed several improvements to the semiclassical description
of BECs in three dimensions.  First, an ansatz that makes it possible
to reduce the corresponding 3D GPE to an effectively 1D equation in
the cylindrically symmetric case was put forward.  An interesting
feature of this approach is that the corresponding 1D normalization
condition, which follows from the standard normalization condition in
the full 3D description, takes a non-canonical form, containing the
fourth power of the 1D wave function, rather than its square.  Also
non-canonical is an expression for the probability density of the
distribution of the 1D momentum.  These results were further extended
to cases when the BEC has vorticity, and when cylindrical symmetry is
absent; these cases yield additional examples of non-canonical
normalization conditions, sometimes of quite complicated form.
 
Another result, obtained in the framework of the effectively 1D
description, is an explicit matching formula between the TF
approximation valid in the classically allowed region, and the
exponentially vanishing WKB approximation valid in the classically
forbidden region.  Here, an exact solution to a problem, found long
ago in an abstract mathematical context, determines the arbitrary
constant in front of the exponentially decaying WKB wave function.
 
To verify the validity of analytical approximations developed in this
work, we have performed direct numerical calculations of bound states
and of dynamics in a time-depenendent potential, and compared the
probability distributions obtained with full 3D results.  The
comparison shows that the analytical approximations are quite
accurate.
 
\begin{acknowledgments} 
We thank M.J. Ablowitz for a useful discussion.  This work was 
supported in a part by grants No.~1998-421 and 1999-459 
from the U.S.-Israel Binational Science 
Foundation, the Israel Science 
Foundation (grant No.~212/01) and the Israel Ministry of Defense 
Research and Technology Unit. 
\end{acknowledgments}

\begin{figure}[!htb] 
\centerline{ 
\includegraphics[width=3in,angle=270,keepaspectratio]{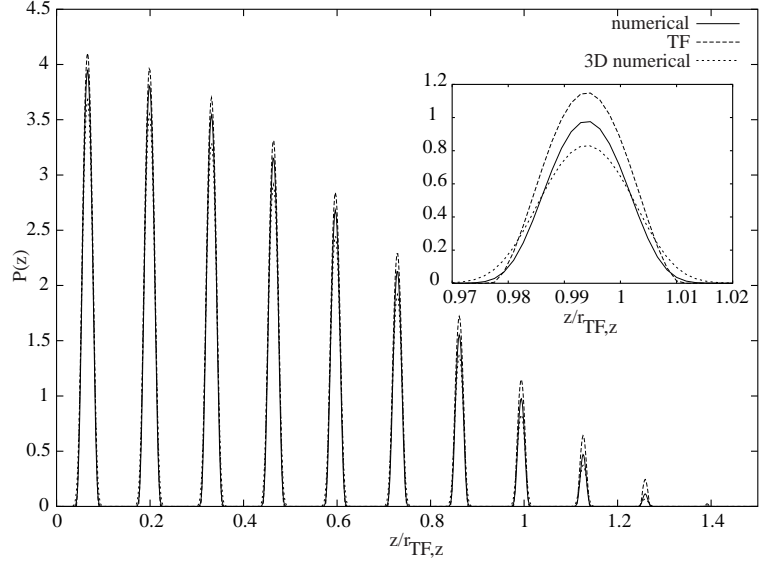}}  
\caption{The probability distributions $P(z)$, $P_{3D}(z)$ and $P_{\mathrm{\ 
TF }}(z)$ versus $z/r_{\mathrm{TF},z}$ for the combination of the 3D 
harmonic and optical potentials. The inset is a blowup of the region near  
$z/r_{\mathrm{TF},z} = 1$.} 
\label{fig3} 
\end{figure} 
 
\begin{figure}[!htb] 
\centerline{ 
\includegraphics[width=3in,angle=0,keepaspectratio]{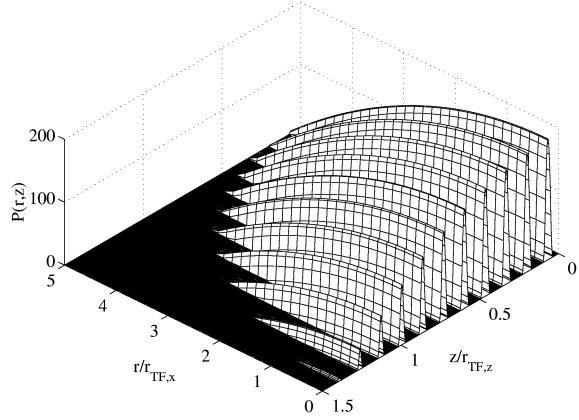}}  
\caption{The probability $|\Psi(\mathbf{r})|^2$ vs $x$ and $z$.} 
\label{fig4} 
\end{figure} 
 
\begin{figure}[!htb] 
\centerline{ 
\includegraphics[width=3in,angle=270,keepaspectratio]{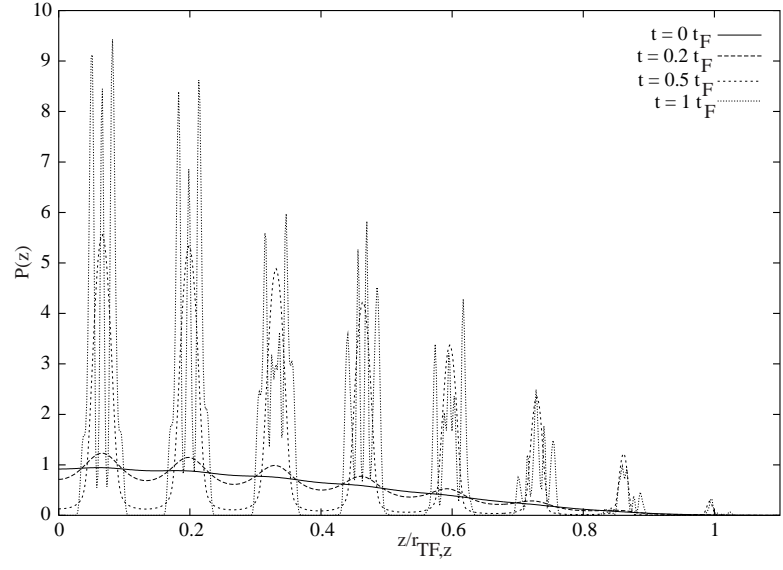}}  
\caption{The probability $P(z,t)$ vs $z$ at four different values of 
time, $t = 0$ [$V_0(0) = 0$], $t=t_F/5$ [$V_0(t_F/5) = 0.387 \, E_R$],  
$t=t_F/2$ [$V_0(t_F/2) = 1.839 \, E_R$], $t=t_F$ [$V_0(t_F) = 5 \, E_R$].} 
\label{fig5} 
\end{figure} 
 
\end{document}